\documentstyle[aps,prl,floats,epsfig,twocolumn,amssymb]{revtex}
\begin{document}
\draft
\wideabs{
\title{Magnetic resonance in the antiferromagnetic and normal state of NH$_{3}$K$_{3}$C$_{60}$}
\author{Ferenc\ Simon$^{1}$, Andr\'{a}s J\'{a}nossy$^{1,}$\cite{email}, Ferenc Mur\'{a}nyi$^{1}$,
Titusz Feh\'{e}r$^{1}$,\\ 	  
Hideo Shimoda$^{2,}$\cite{shim}, Yoshihiro Iwasa$^{2}$, and
L\'{a}szl\'{o} Forr\'{o}$^{3}$}
\address{$^{1}$Technical University of Budapest, Institute of Physics, H-1521\\
Budapest, PO BOX 91, Hungary\\
$^{2}$Japan Advanced Institute of Science and Technology, Tatsunokuchi,\\
Ishikawa 923-1292, Japan\\
$^{3}$Laboratoire de Physique des Solides Semicristallins, IGA Department de\\
Physique, Ecole Polytechnique Federal de Lausanne, 1015 Lausanne, Switzerland}
\date{\today}
\maketitle

\begin{abstract}
We report on the magnetic resonance of NH$_{3}$K$_{3}$C$_{60}$ powders in the
frequency range of 9 to 225 GHz. The observation of an antiferromagnetic
resonance below the phase transition at 40 K is evidence for an
antiferromagnetically ordered ground state. In the normal state, above 40 K,
the temperature dependence of the spin-susceptibilty measured by ESR agrees
with previous static measurements and is too weak to be explained by
interacting localized spins in an insulator. The magnetic resonance
line width has an unusual magnetic-field dependence which is large and
temperature independent in the magnetically ordered state and decreases
rapidly above the transition. These observations agree with the suggestion
that NH$_{3}$K$_{3}$C$_{60}$ is a metal in the normal state and undergoes a
Mott-Hubbard metal to insulator transition at 40 K.
\end{abstract}

\pacs{71.30.+h,74.70,76.50}
}
The superconducting transition temperature, $T_{c}$, of cubic alkali metal A$%
_{3}$C$_{60}$ fullerides has a simple relation to the lattice constant 
\cite{gunnarssonrmp97}: $T_{c}$ increases with the ionic size of the alkali.
This motivates the quest to synthesize large lattice parameter fullerides
with $C_{60}$ in the $($C$_{60})^{3-}$ charge state. The intercalation of
ammonia into Na$_{2}$CsC$_{60}$ increased $T_{c}$ by nearly 20 K, from 10.5
K\ to 29.7 K\cite{zhounature93}. In contrast, superconductivity was not
detected at ambient pressure in NH$_{3}$K$_{3}$C$_{60}$, the ammonia
intercalated stable phase of K$_{3}$C$_{60}$\cite{rosseinskynature93}.
According to Raman scattering in NH$_{3}$K$_{3}$C$_{60}$ the charge state is 
$\ ($C$_{60})^{3-}$ and, at least along some directions, the
nearest-neighbor interball separation exceeds substantially that of Rb$_{2}$%
CsC$_{60}$, a superconductor with one of the highest transition temperatures\cite{zhouprb95}. The two materials differ in that NH$_{3}$ intercalation expands the
lattice without changing the cubic symmetry in Na$_{2}$CsC$_{60}$ while NH$%
_{3}$K$_{3}$C$_{60}$ has a face-centered orthorhombic structure. Zhou et al.
questioned the importance of lower symmetry since superconductivity is
recovered under hydrostatic pressure at a relatively high temperature (28 K
at 14.8 kbar) without any change in the crystallographic structure\cite
{zhouprb95}. They suggested that the smaller transfer integrals at ambient
pressures suppresses superconductivity and favors a Mott-Hubbard transition
to an insulating ground state. The observation by Iwasa et al.\cite
{iwasaprb96} and by Allen et al.\cite{allenjmatchem96} of a phase transition
with no structural change\cite{ishiiprb99} at about 40 K\ and ambient
pressure reinforces this view. The ESR intensity at 9 GHz disappeared below
40 K but the static susceptibility, $\chi $, measured by SQUID, and the $%
^{13}$C\ NMR\cite{iwasaprb96} changed little. In spite of these somewhat
contradicting results, Iwasa et al. \cite{iwasaprb96} concluded that the
ground state is an antiferromagnet. Initially it was thought that the
ordered magnetic moment is unusually small, $\sim 0.01$ $\mu _{B}$.
Recently, Tou et al.\cite{touphysb99} found that the $^{13}$C line is significantly broadened below the transition, indicating an
antiferromagnetically ordered ground state of NH$_{3}$K$_{3}$C$_{60}$ with
an ordered moment of $1$ $\mu _{B}/$C$_{60}$. In the interpretation of the $%
^{13}$C NMR\cite{touphysb99} it was assumed that spins lie perpendicularly
to the external field and the spin-flop(SF) field is small, less than 1 T.
The recent detection of damped oscillations in zero-field $\mu $SR \cite
{prassidesphysc97} is a further signature of static magnetic moments below
40 K. Just like the ground state, the normal state of NH$_{3}$K$_{3}$C$_{60}$
is also poorly understood. In particular, it is unknown whether above 40 K
it is a metallic or insulating. Initially, the temperature variation of $%
\chi $ and $^{13}$C NMR $T_{1}$ of \cite{iwasaprb96} were interpreted
assuming a narrow band metallic normal state. Recent NMR\ data\cite
{touphysb99} were analyzed assuming localized spins on the C$_{60}$ ions
above 170 K. A further complication arises from the recently found
structural transition \cite{ishiiprb99} due to the ordering of the K$-$NH$%
_{3}$ pairs at $T_{S}=150$ K.

The motivation for the present work is two-fold: following Ref.\cite
{janossyprl97} on RbC$_{60}$ and CsC$_{60}$ we observe the antiferromagnetic
resonance(AFMR) in NH$_{3}$K$_{3}$C$_{60}$ using a multi-frequency spin
resonance technique. Our less detailed earlier work\cite{simon98} was
inconclusive in this respect. Below 40 K the 9 GHz resonance\ spectrum of NH$%
_{3}$K$_{3}$C$_{60}$ is broad and good sensitivity and high purity material
is required for its detection\cite{simon98}. The AFMR in powders becomes
narrower at higher fields and is an unambiguous evidence for the
antiferromagnetic(AF) ground state. Secondly, we gain information about the
normal state from the ESR measured in a large frequency range. Our normal
state ESR susceptibility data agree with the static susceptibility data of
Ref.\cite{iwasaprb96} and point to a metallic system rather than an
insulator with interacting localized moments.

Several NH$_{3}$K$_{3}$C$_{60}$ powder samples were prepared following
methods described elsewhere\cite{iwasaprb96}. Samples were sealed in quartz
ESR tubes under low pressure He. Powder X-ray diffraction showed them to be
of high purity. The small residual K$_{3}$C$_{60}$ content was unobservable
by X-ray diffraction but increased the noise of the 9 GHz ESR\ spectrometer
below $19$ K. The main features of the ESR spectra were identical for all
samples studied and we quote in our figures results from two samples. 9 GHz
ESR spectra were recorded on a commercial Bruker spectrometer. The 35, 75,
150 and 225 GHz ESR\ were studied at the Budapest high frequency
spectrometer. Samples were slowly ($\sim $50 K/h) cooled from 300 K to 5
K. At 9 GHz only data between 20 and 300 K\ are shown, as below 19 K residual K$%
_{3}$C$_{60}$ distorted the spectra. ESR intensities and g-factors were
measured with respect to standard calibrating samples.


In Fig. 1 we show $\chi $ determined from the ESR intensity together with $%
\chi $ measured in Ref.\cite{iwasaprb96} by SQUID technique. $\chi $ of NH$%
_{3}$K$_{3}$C$_{60}$ varies little through the 40 K\ transition for all ESR
frequencies, $f$, between 9 (0.3 T)\ and 225 GHz (8.1 T). Our $\chi $ data
are in remarkable agreement with that of the SQUID results of Iwasa et al. 
\cite{iwasaprb96} in the full temperature range. The SQUID measurements were
corrected for the core electron contribution. ESR measures the spin
susceptibility in the paramagnetic state and, if the applied field is large
compared to the SF field, in the AF state also. Only the perpendicular AF\
susceptibility is observed in such large fields. The (molar) susceptibility
below $T_{N}$ is $\chi _{\perp }=N_{A}\frac{\mu _{\text{eff}}^{2}}{3k_{B}}%
\frac{1}{2T_{N}}$, where $N_{A}$ is the Avogrado number. From Fig. 1 $\chi
_{\perp }\simeq 1.3\ast 10^{-3}$ emu/mol and the ordered moment is $\mu _{%
\text{eff}}=0.9$ $\mu _{B}/$C$_{60}$.

The variation of $\chi (T)$ with $T$ is difficult to explain with a
''textbook'' metallic or localized spin susceptibility. The variation is too
weak to originate from a Curie-Weiss temperature dependence of localized
spins. The 3D AF order sets in at a Ne\'{e}l temperature of $T_{N}\simeq 40$
K. Between 40 and 300 K\ $\chi (T)$ decreases by a factor $\sim 2.3$ whereas
a factor 4.25 would be expected from a Curie-Weiss behavior of
antiferromagnetically correlated spins $\chi (T)=\frac{C}{T+T_{N}}$, where C
is the Curie constant. The fit of $\chi (T)$ to a Curie-Weiss behavior with $%
T_{N}$ left as a free parameter leads to $T_{N}\geqq 120$ K in contradiction
with the observed long range 3D order at 40 K. On the other hand, the
variation of $\chi $ is larger than usually found in metallic systems.
Nevertheless, strong correlation effects in a narrow band metal may lead to
the observed variation of $\chi (T)$ as it was originally suggested in Ref. 
\cite{iwasaprb96}. Electron-electron correlation effects are needed to
explain the large value of $\chi $ of the K$_{3}$C$_{60}$ parent compound 
\cite{gunnarssonrmp97} also. The larger lattice constants of NH$_{3}$K$_{3}$C%
$_{60}$ suggest a narrower band and even more important correlation effects.
There is a slight anomaly in $\chi (T)$ at the structural transition, $T_{S}$%
. The transition is of second order and affects $\chi $ in the range of
50-150 K. The sensitivity of $\chi (T)$ to the structural transformation is
readily understood if the material is a metal with strong electron
correlations since the freezing of molecular rotation is certainly
accompanied by a change in electronic structure. In summary, above $T_{N}$
the material is more likely a strongly correlated metal than an insulator
with localized spins. Unfortunately, a definite answer is not possible from
the $\chi $ data alone.


Fig. 2 shows the variation of the resonance line width ($\Delta H$) with
temperature at various frequencies. The 40 K transition is marked by a
sharp increase of $\Delta H$ with decreasing $T$. The 9, 35, and 75 GHz
spectra below 40 K are characteristic of an AFMR of a powder in applied
fields larger than the SF field\cite{janossyprl97}: the line width narrows
and the resonance field shifts proportionally to the inverse of the applied
field. This is in contrast with paramagnetic systems where (excluding crystal
field effects) $\Delta H$ is either independent of $f$ or increases with $f$%
. We rule out a spin-glass ground state, proposed in Ref.\cite{mehring} for RbC$_{60}$, since we observed no thermal or magnetic history dependence. Also, were
the ground state a spin-glass, there would be no simple explanation for the
scaling of the line shape. As shown in Fig. 3, the 24 K line shapes are
identical at 9 and 35 GHz if the magnetic field axis is multiplied by the resonance field, $H_{0}$. The close similarity
of the scaled line shapes is a characteristic feature of AFMR in powders and
implies a $1/H_{0}$ $(1/f)$ dependence of $\Delta H$ below 40 K. As
expected, below $T=40$ K the resonance field shift (not shown), $H_{0}(T)-H_{0}(40$ K$)$ scales also with $1/H_{0}$ at 9 and 35 GHz.


Between 75 and 225 GHz the line width depends little on $f$ and does not
follow the $1/H_{0}$ field dependence (Fig. 4.).\ \ To explain this, we
assume that in this range the $1/f$ narrowing of the AFMR\ line is
compensated by a g-factor anisotropy broadening linear in $f$. The g-factor
anisotropy is a consequence of the orthorhombic crystal structure of NH$_{3}$%
K$_{3}$C$_{60}$\cite{rosseinskynature93}. For simplicity, we assume in the
following a uniaxial anisotropy and that the principal axes of the g-tensor
coincide with the easy and hard axes of the AF state. In a single crystal
the resonance field depends on the alignment of the external field with
respect to the magnetic axes of the antiferromagnet. The hard and easy\ axis
modes correspond to external fields perpendicular and parallel\ to the easy
axis, respectively\cite{foner}. If $H_{0}$ is larger than the SF field, $%
H_{SF}$, the modes are\cite{foner} $\omega _{\pm }=\gamma _{\pm }\sqrt{%
H_{0}^{2}\pm H_{SF}^{2}}$. The +(-) sign corresponds to the hard (easy)\ axis mode, $\gamma _{\pm }=\gamma \pm \Delta \gamma /2$ where $\gamma =g\mu _{B}/\hbar $, $\Delta \gamma $ measures the anisotropy. In a powder of independent single crystals the spectrum has extrema at $H_{AFMR}^{\pm }=\sqrt{\omega _{{}}^{2}/\gamma _{\pm }^{2}\mp H_{SF}^{2}}$, ($\omega =2\pi f)$. We assume that the measured linewidth $\Delta H=$ $H_{AFMR}^{-}-H_{AFMR}^{+}$.
A fit to this formula (Fig. 4) with the free parameters $%
H_{SF}$ and $\Delta \gamma $ is in good agreement with experiment. We find $%
\Delta \gamma =1300\pm 200$ ppm independent of $T$ below 40 K and $H_{SF}(5$
K$)=0.17$ $T$. The anisotropy in NH$_{3}$K$_{3}$C$_{60}$ is large in
comparison with $\Delta \gamma /\gamma \simeq 80$ ppm in the orthorhombic
fulleride polymer $Rb$C$_{60}$ at $T_{N}$. On the other hand, $H_{SF}$ in NH$%
_{3}$K$_{3}$C$_{60}$\ is close to values in the $Rb$C$_{60}$ and $Cs$C$_{60}$
fulleride antiferromagnets \cite{janossyprl97} The SF field is $%
H_{SF}=(2H_{E}H_{A})^{1/2}$ where $H_{E}$ and $H_{A}$ are the exchange and
anisotropy fields, respectively. NH$_{3}$K$_{3}$C$_{60}$ is composed of
light elements where spin-orbit interactions are small and $H_{A}$ arises
from dipolar fields and is of the order of 1 mT. Since $g\mu _{B}H_{E}$ is
of the order of $kT_{N}$, we expect $H_{SF}$ is of the order of 0.1-1 T.


The assignment of the resonance below 40 K to an AFMR powder spectrum with a
g-factor anisotropy implies that the broadening is inhomogeneous. However,
although the spectra are significantly broadened, the line shapes are
Lorentzian, which is characteristic of a homogeneous relaxational
broadening. We do not know of any relaxational mechanism that could explain
the data. It may be that the interaction at the boundaries of small domains
with differing crystal orientations reduces the inhomogeneous broadening of
the AFMR and renders the lineshape approximately Lorentzian. In this case
the values of $H_{SF}$ and $\Delta \gamma /\gamma $ are only lower limits.


The ESR in the normal state (above 40 K)\ has an interesting frequency and
temperature dependence (Fig. 2, inset and Fig. 5). The line width, $\Delta H,$
can be decomposed into $\Delta H(f,T)=\Delta H_{0}(T)+\Delta H_{f}(T)$, where $\Delta H_{f}(T)$ is proportional to $f$. The frequency independent term $\Delta H_{0}(T)$ has a maximum at the structural
transition, $T_{S}$. This maximum is observed in the 9 GHz line width but is
hidden at higher frequencies by the frequency dependent term (Fig. 2, inset).
As the transition is approached from above, molecular rotations of the NH$%
_{3}$ groups become slower and fluctuations at the Larmor frequency broaden
the line . Below the transition the molecular order increases gradually and
the line narrows. A static or slowly fluctuating disorder is the usual
mechanism for conduction electron spin relaxation and a maximum is expected if the state is metallic.

An incompletely resolved g-factor anisotropy is the most likely origin of \ $%
\Delta H_{f}(T)$ \ in orthorhombic NH$_{3}$K$_{3}$C$_{60}$. ($\Delta H$ is
independent of frequency in \ cubic K$_{3}$C$_{60}$). $\Delta
H_{f}(T)=H_{0}$ $\Delta \gamma /\gamma $ is proportional to $f$ up to 225
GHz at and above 40 K. As shown in Fig. 5, $\Delta \gamma /\gamma $ increases
strongly as the temperature approaches $T_{N}$ from above. The g-factor
anisotropy at 40 K, (i.e. slightly above $T_{N}$) is equal to the
temperature independent value in the AF state. The strong non-linearity in
the frequency dependence of the linewidth appears suddenly at temperatures
slightly below 40 K. Thus fields up to 8 T do not smear the transition. Such
behavior may be expected for an AF ordering with a large value of $H_{E}$
but is unlikely in a spin glass where the melting temperature is smeared by
magnetic fields. The g-factor broadening must be incomplete in the
paramagnetic state for the same reason as in the AF state. Although at 45 K
the ESR is several times broader at 225 GHz than at 9 GHz the line shape is
Lorentzian in both cases. The ESR of high purity Al is an example for an
incomplete g-factor anisotropic broadening \cite{lubzensprl72} in a metal.
In NH$_{3}$K$_{3}$C$_{60}$, the increase of the g-factor anisotropy may be
due to a gradual metal to insulator transition where the motional narrowing
due to the diffusion of conduction electrons becomes less effective as $T_{N}
$ is approached.

The observation of an AFMR below 40 K is an unambiguous evidence for an antiferromagnetically ordered ground state. The static susceptibility and ESR are compatible with a paramagnetic metallic state at high temperature. This could happen if\ NH$_{3}$K$_{3}$C$_{60}$
undergoes a Mott-Hubbard metal to insulator transition. If this suggestion
is true then NH$_{3}$K$_{3}$C$_{60}$ is one of the rare examples, and would
be the only example among fulleride compounds, in which a Mott-Hubbard
transition takes place at experimentally accessible temperatures.

Support from the JSPS-HAS Japanese-Hungarian Cooperation Program,
the Hungarian State Grants OTKA T029150, FKFP0352-1997, and HAS-TUB No. 04119
on Solids in Magnetic Fields, JSPS (RFTF96P00104), the Ministry of Education, Science, Sports, and Culture of Japan, and the Swiss NSF are acknowledged.

\pagebreak
Figure Captions:

Fig. 1. Spin-susceptibility of NH$_{3}$K$_{3}$C$_{60}$ measured by ESR at 9 GHz
($\blacksquare $ 0.3 T) and 225 GHz ($\diamond $ 8.1 T). Solid curve is
SQUID data from Ref.\protect\cite{iwasaprb96} measured at 1 T after core
electron diamagnetic susceptibility correction. $T_{S}$
is the structural, $T_{N}$ the magnetic ordering temperature.

Fig. 2. Temperature dependence of the line width, $\Delta H$, at various
frequencies. ($\blacksquare $ 9 GHz, $\bigcirc $ 35 GHz, $\blacktriangle $
75 GHz) Inset shows $\Delta H$ (in mT) above 40 K.

Fig. 3. Antiferromagnetic resonance at 24 K. Line shapes at 9 and 35 GHz
are identical if the magnetic field scale is multiplied by the resonance
field, $H_{0}$. (Low intensity narrow impurity lines are subtracted from the
spectra.) We chose the normalizing constant $K=0.17$ $T$, the spin flop
field $H_{SF}$ measured at 5 K.

Fig. 4. $\Delta H$ as a function of $f$ at various temperatures. The
narrowing of the AFMR is compensated at high frequencies by a g-factor
anisotropy broadening. (Solid lines: fit, explained in text.) Inset:
$\Delta H$ (in mT) is linear in $f$ due to partially resolved g-factor
anisotropy. (Dashed lines: linear fit.)

Fig. 5. Temperature variation of normalized linearly $f$ dependent
contribution. (Dashed line indicates $T_{N}$.) $\Delta \protect\gamma /%
\protect\gamma $ is assumed to be due to partially resolved g-factor
anisotropy in both the AF and paramagnetic state.

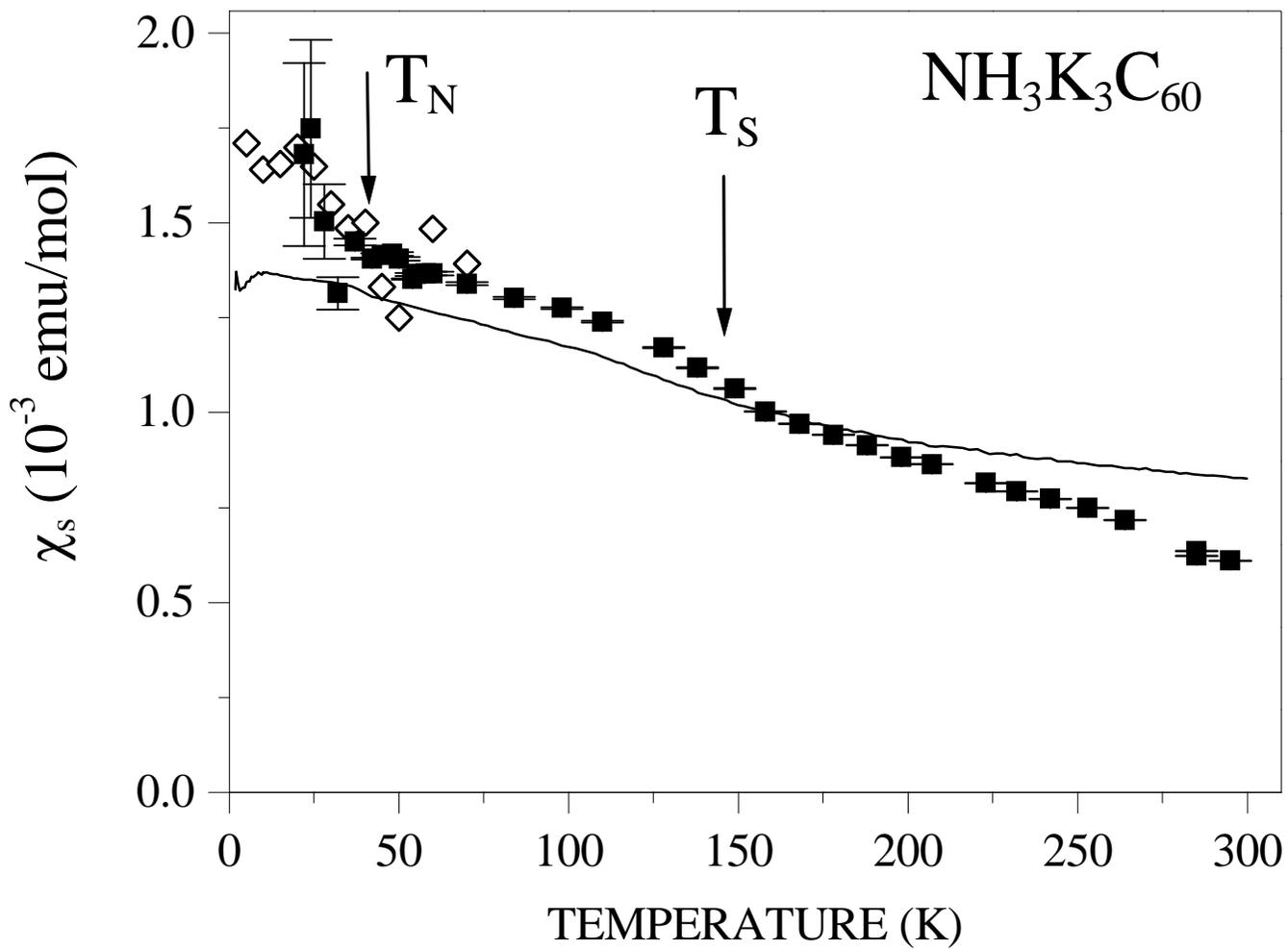
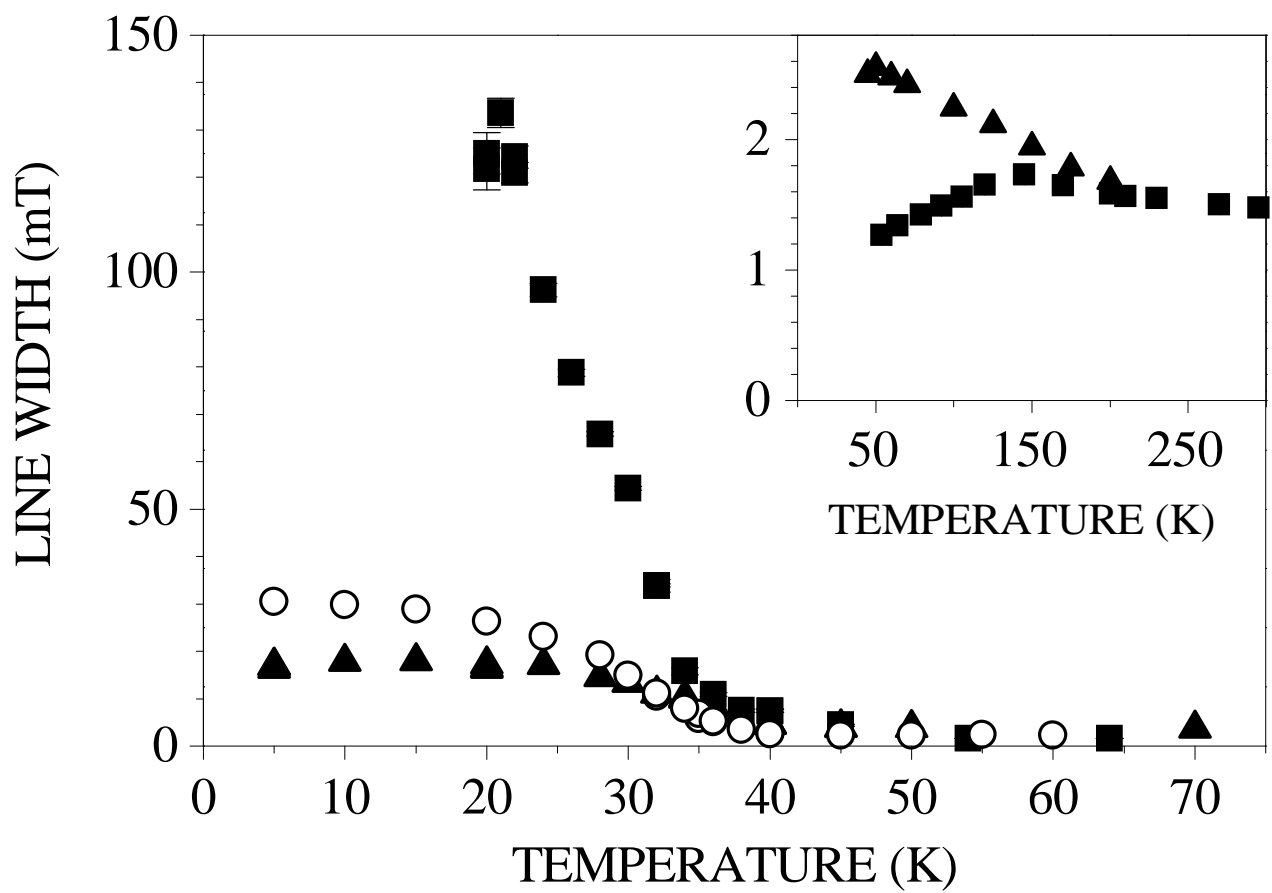
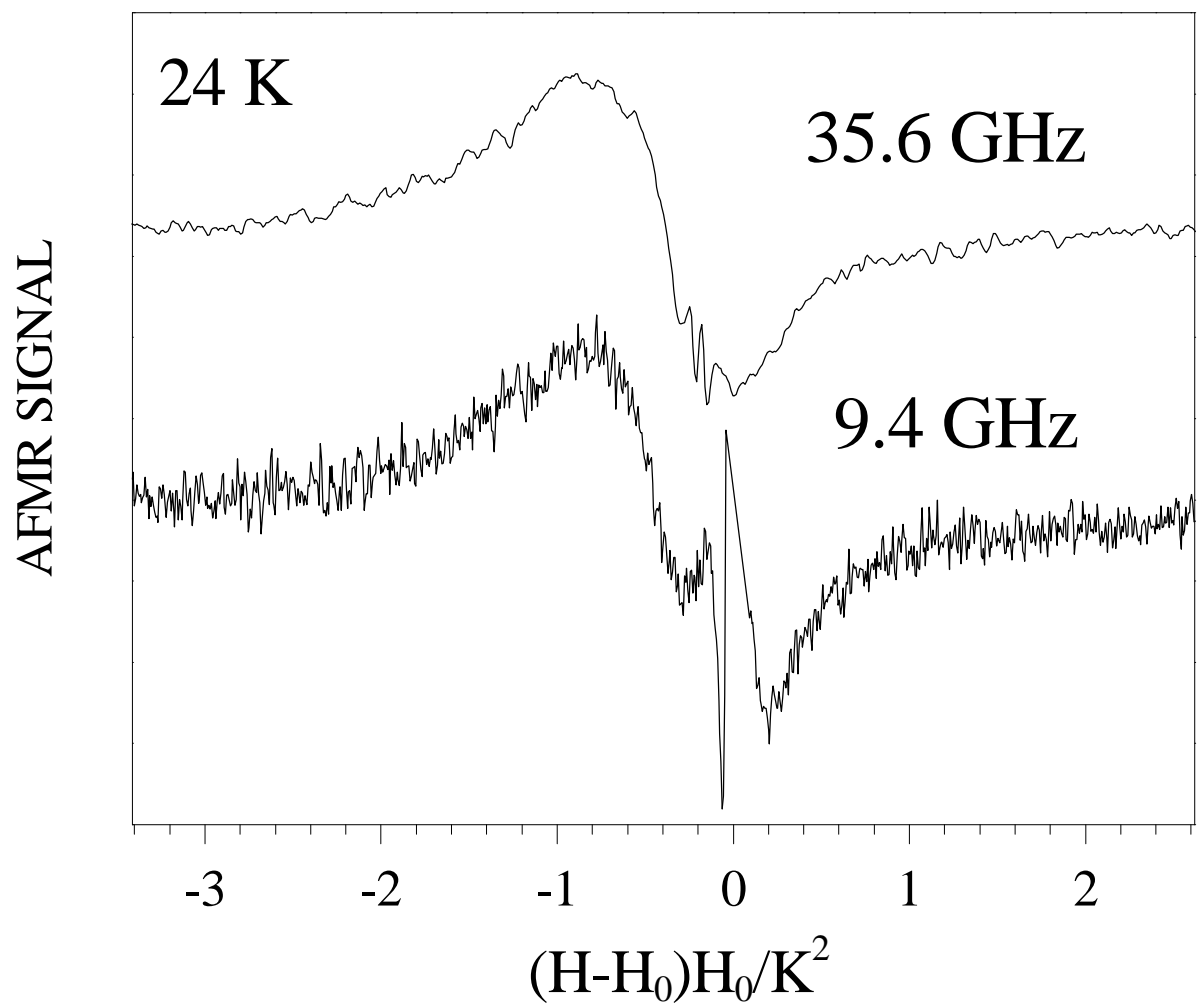
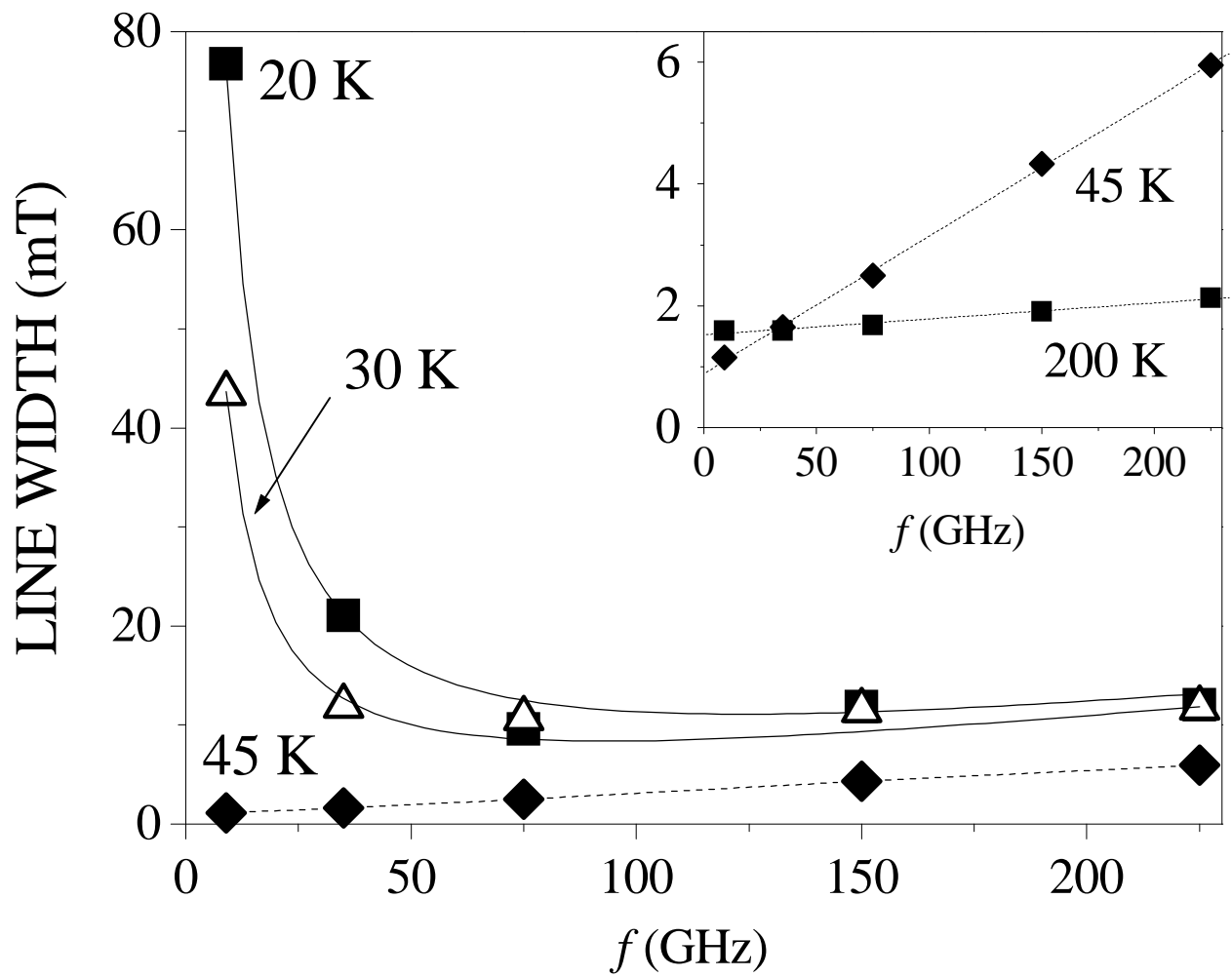
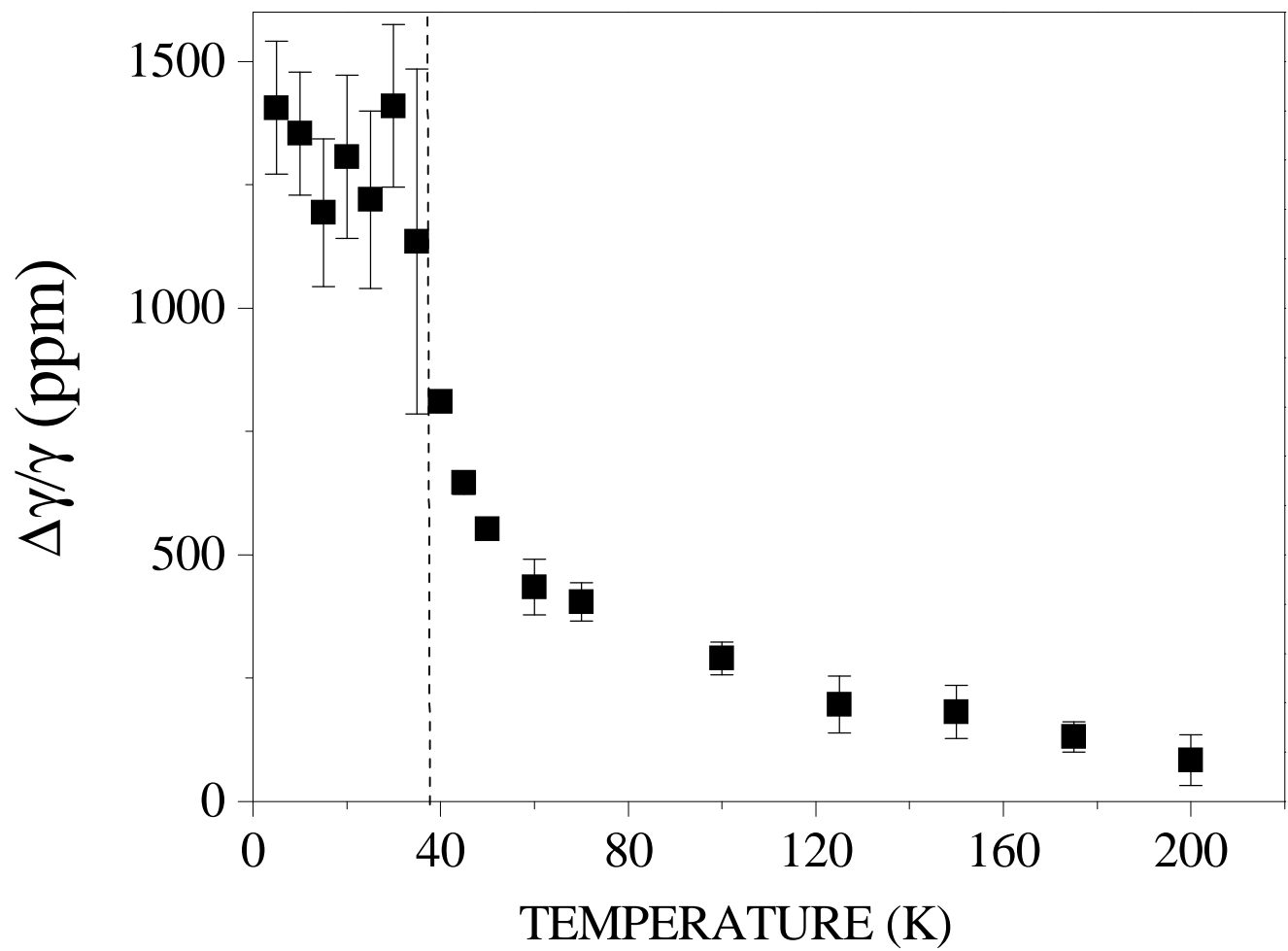
\end{document}